\documentstyle[aps,prb,epsf,twocolumn,floats]{revtex}

\begin{document}
\draft
\title{Scattering approach to parametric pumping}

\author{P. W. Brouwer}
\address{Lyman Laboratory of Physics, Harvard University, 
  Cambridge, MA 02138, USA\\
{\rm (\today)}
\medskip ~~\\ \parbox{14cm}{\rm
A d.c.\ current can be pumped through a quantum dot by periodically
varying two independent parameters $X_1$ and $X_2$, like a gate voltage
or magnetic field. We present a formula that relates the pumped current
to the parametric derivatives of the scattering matrix $S(X_1,X_2)$ of
the system. As an application we compute the statistical
distribution of the pumped current in the case of a chaotic quantum
dot.\medskip
}}
\maketitle 

\narrowtext

An electron pump is a device that generates a d.c.\ current between two
electrodes that are kept at the same bias. In recent years, electron
pumps consisting of small semiconductor quantum dots have received
considerable experimental and theoretical 
attention.\cite{KJVHF,PLUED,BS,SZB,SN,SS,OKKVH,BBS,F,BP,AA} 
A quantum dot is a
small metal or semiconductor island, confined by gates, and connected
to the outside world via point contacts.  Several different mechanisms
have been proposed to pump charge through such systems, ranging from a
low-frequency modulation of gate voltages in combination with the
Coulomb blockade \cite{KJVHF,PLUED,AA} to photon-assisted transport at or
near a resonance frequency of the dot.\cite{SN,SS,OKKVH,BBS} Their
applicability depends on the characteristic size of the system and the
operation frequency.

Most experimental realizations of electron pumps in semiconductor
quantum dots made use of the principle of Coulomb blockade. If the dot
is coupled to the outside world via tunneling point contacts, the
charge on the dot is quantized, and (apart from degeneracy points)
transport is inhibited as a result of the high energy cost of adding an
extra electron to the dot.  
Pothier et al.\ constructed an electron pump that operates at
arbitrarily low frequency and with a reversible pumping 
direction.\cite{PLUED} 
The pump consists of two weakly coupled quantum dots in
the Coulomb blockade regime and operates via a mechanism that closely
resembles a peristaltic pump: Charge is pumped through the double dot
array from the left to the right and electron-by-electron as the
voltage $U_1 \propto \sin(\omega t)$ of the left dot reaches its minima
and maxima before the voltage $U_2 \propto \sin(\omega t - \phi)$ of
the right one.\cite{PLUED} The pumping direction can be reversed by
reversing the phase difference $\phi$ of the two gate voltages.

\begin{figure}
  \epsfxsize=0.8\hsize
  \hspace{0.09\hsize}
  \epsffile{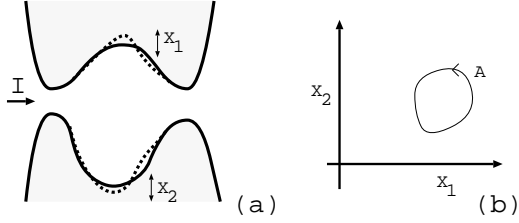}\\

\caption{ \label{fig:1} (a): A quantum dot with
two parameters $X_1$ and $X_2$ that describe a deformation of the shape
of the quantum dot. As $X_1$ and $X_2$ are varied periodically, a d.c.\
current $I$ is generated. (b) In one period, the parameters $X_1(t)$ and $X_2(t)$ follow a closed path in parameter space. The pumped current
depends on the enclosed area $A$ in $(X_1,X_2)$-parameter space.}
\end{figure}

A similar mechanism was proposed by Spivak, Zhou, and Beal Monod for an
electron pump consisting of single quantum dot only.\cite{SZB} In this
case a d.c.\ current is generated by adiabatic variation of two
different gate voltages that determine the shape of the nanostructure,
or any other pair of parameters $X_1$ and $X_2$, like magnetic field or
Fermi energy, that modify the (quantummechanical) properties of the
system, see Fig.\ \ref{fig:1}a. The magnitude of the current is
proportional to the frequency $\omega$ with which $X_1$ and $X_2$ are
varied and (for small variations) to the product of the amplitudes
$\delta X_1$ and $\delta X_2$. The direction of the current depends on
microscopic (quantum) properties of the system, and need not be known a
priori from its macroscopic properties. As in the case of the
double-dot Coulomb blockade electron pump of Ref.\ \onlinecite{PLUED},
the direction of the current in the single-dot parametric pump of
Spivak et al.\cite{SZB}\ can be reversed by reversing the phases of
the parameters $X_1$ and $X_2$. 
An important difference between the two mechanisms is that a parametric
electron pump like the one of Ref.\ \onlinecite{SZB} does not require
that the quantum dot is in the regime of Coulomb blockade; it operates
if the dot is open, i.e.\ well coupled to the leads by means of
ballistic point contacts.
Experimentally, an electron pump in an open quantum dot has been
realized only very recently.\cite{Marcus} A measurement of the pumped
current provides a promising tool to study properties of open
mesoscopic systems at zero bias or at zero current.

In this paper we consider a parametric electron pump through an open
system in a scattering approach. 
Our main result is a formula for the pumped current in terms of the scattering
matrix $S(X_1,X_2)$.
Such a formula
is the analogue of the Landauer formula, which relates the conductance
$G = \delta I/ \delta V$ of a mesoscopic system with two contacts to a sum over the (squares of) matrix elements $S_{\alpha \beta}$,
\begin{equation} \label{eq:Landauer}
  \delta I = G\, \delta V = {2 e^2 \over h} \delta V \sum_{\alpha \in 1} \sum_{\beta \in 2} |S_{\alpha \beta}|^2.
\end{equation}
The indices $\alpha$ and $\beta$ are summed over all channels in
the left and right contacts, respectively, and $\delta V$ is the
applied voltage. For the case of the parametric electron
pump, where two parameters $X_1$ and $X_2$ are varied periodically,
$\delta X_1(t) = \delta X_1 \sin(\omega t)$ and $\delta X_2(t) = \delta
X_2 \sin(\omega t - \phi)$, we find that the d.c.\ component of the
current $I$ depends on the {\em derivatives} $\partial S_{\alpha
\beta}/\partial X$,
\begin{eqnarray} \label{eq:PumpFirst}
  \delta I &=& {\omega e \sin \phi\, \delta X_1 \delta X_2 \over 2 \pi}
    \sum_{\alpha\in 1} \sum_{\beta} \mbox{Im}\,
    {\partial S_{\alpha \beta}^{*} \over \partial X_1}
    {\partial S_{\alpha \beta}^{\ } \over \partial X_2}.
\end{eqnarray}
Like the Landauer formula, Eq.\ (\ref{eq:PumpFirst}) is valid for
a phase coherent system at zero temperature and to
(bi)linear response in the amplitudes $\delta X_1$ and $\delta X_2$.
[The nonlinear response is given by Eq.\ (\ref{eq:Pump}) below.] It
captures both a classical contribution to the current and the 
quantum interference corrections. Quantum corrections can be important
in the mesoscopic regime, especially if
there is no ``classical'' mechanism that dominates the pumping 
process.\cite{SZB,Marcus}
Eq.\ (\ref{eq:PumpFirst}) is valid to first order in the frequency
$\omega$.  This is sufficient if the period $\tau = 2 \pi/\omega$ is
much larger than the time particles spend inside the quantum dot. For
such low frequencies, we can assume that equilibrium is maintained
throughout the pumping process. The scattering matrix formula does not
capture effects of order $\omega^2$ (or higher) that rely on the
existence of a non-equilibrium distribution inside the quantum dot.
\cite{SZB}  The existence of a scattering approach to parametric
pumping allows us to borrow from the vast literature dealing with
scattering matrices of disordered and chaotic microstructures and their
parameter dependence,\cite{Review} and to directly relate the pumped
current to other transport properties like e.g.\ the conductance.

The system under consideration is shown schematically in
Fig.\ \ref{fig:1}a. It consists of a quantum dot, coupled to two
electron reservoirs by ballistic point contacts. The two electron
reservoirs are held at the same voltage. Two external parameters
$X_1(t)$ and $X_2(t)$ of the dot are varied periodically, see
Fig.\ \ref{fig:1}a. They can be e.g.\ the voltage of a plunger
gate, parameters that characterize the shape, or a magnetic field.
The two point contacts, which have $N$ channels at the Fermi level
$E_F$, are labeled $1$ and $2$. The scattering matrix $S$ of the system
has dimension $2N \times 2N$ and is a function of the parameters $X_1$
and $X_2$. Since the system is well coupled to the leads, the charge is
no longer quantized, the Coulomb blockade is lifted, and to a first
approximation, we can use a picture of non-interacting 
electrons.\cite{footnote}

Starting point of our theory is a formula due to B\"uttiker, Thomas,
and Pr\^etre \cite{BTP} for the current in the contacts $1$ and $2$ 
that results from an infinitesimal change of a parameter $X$: For a small
and slow harmonic variation $X(t) = X_0 + \delta X_{\omega} e^{i \omega
t}$, the charge $\delta Q(m)$ entering the cavity through contact
$m$ ($m=1,2$) reads
\begin{mathletters}\label{eq:dQ}
\begin{eqnarray} 
  \delta Q(m,\omega) &=& e {d n(m) \over d X} \delta X_{\omega}, \\
  {d n(m) \over d X} &=& {1 \over 2 \pi} \sum_{\beta}
  \sum_{\alpha \in m} \mbox{Im}\, {\partial S_{\alpha \beta} \over 
    \partial X} S_{\alpha \beta}^{*}.  \label{eq:emissivity}
\end{eqnarray} 
\end{mathletters}
The index $\alpha$ is summed
from $1$ to $N$ for contact $1$ and from $N+1$ to $2N$ for contact
$2$. The quantity $dn(m)/dX$ is the {\em emissivity} into contact 
$m$.\cite{BTP}  Eq.\ (\ref{eq:dQ}) is valid to first order in the
frequency $\omega$ and assumes that the scattering properties follow
the time-dependent potentials instantaneously.
After Fourier transformation one obtains
\begin{equation}
  \delta Q(m,t) = e {d n(m) \over d X} \delta X (t).
\end{equation}
Similarly, for a simultaneous infinitesimal variation of two parameters
$X_1$ and $X_2$, the emitted charge $\delta Q(m,t)$ through contact $m$
is ($m=1,2$)
\begin{equation} \label{eq:dQQ}
  \delta Q(m,t) = e {d n(m) \over d X_1} \delta X_1 (t) +
                  e {d n(m) \over d X_2} \delta X_2 (t) .
\end{equation}

Next, we consider a {\em finite} variation of both parameters $X_1$ and
$X_2$. The total charge emitted through contact $m$ in one period $\tau
= 2 \pi/\omega$ is found from integration of Eq.\ (\ref{eq:dQQ}) to
$X_1$ and $X_2$, bearing in mind that the scattering matrix $S$ and
hence the emissivities $d n(m) / d X_1$ and $d n(m) / d X_2$ are
functions of $X_1$ and $X_2$,
\begin{eqnarray}
  Q(m,\tau) &=& e \int_0^{\tau} dt \left(
    {d n(m) \over d X_1}
  {d X_1 \over d t} 
  + {d n(m) \over d X_2}
 {d X_2 \over d t} \right).
\end{eqnarray} 
In one period, the pair of parameters $X_1(t)$ and $X_2(t)$ follows a
closed path in the $(X_1,X_2)$ parameter space, see Fig.\ \ref{fig:1}b.
The total charge expelled from the dot through contact $m$ can be
rewritten as a surface integral over the area $A$ enclosed by
the path in parameter space using Green's theorem,
$$
  Q(m,\tau) = e \int_A dX_1 dX_2 \left( 
    {\partial \over \partial X_1} {d n(m) \over d X_2}
    - {\partial \over \partial X_2} {d n(m) \over d X_1}
    \right).
$$
Note that the surface area $A$, and hence the transported
charge, vanish, if the parameters $X_1$ and $X_2$ vary in phase, or
with a phase difference $\pi$. The surface area is maximal if their
phases differ by $\pi/2$.
Substitution of Eq.\ (\ref{eq:emissivity}) for the emissivities yields
\begin{eqnarray}
  Q(m,\tau) &=& {e \over \pi} \int_A dX_1 dX_2 \sum_{\beta} \sum_{\alpha \in m}
    \mbox{Im}\, 
           {\partial S_{\alpha \beta}^{*} \over \partial X_1} 
           {\partial S_{\alpha \beta}^{\vphantom{*}} \over \partial X_2}
     . \label{eq:PumpCurrent}
\end{eqnarray}
Hence the d.c.\ current $I_m$ through contact $m$ is given by 
\begin{mathletters} \label{eq:Pump}
\begin{eqnarray}
  I_m &=& {i \omega e \over 4 \pi^2}
    \sum_{\alpha\in m} \int_A dX_1 dX_2
    \left[ R_{X_1}, R_{X_2} \right]_{\alpha \alpha}, \\
  R_X &=& -i {\partial S \over \partial X} S^{\dagger}.
\end{eqnarray}
\end{mathletters}%
One verifies that $I_1 = - I_2$, indicating that no charge is
accumulated. The response matrices $R_{X_1}$ and $R_{X_2}$
are hermitian $2N \times 2N$ matrices. For the
(bi)linear response to the variations of the parameters $X_1$ and
$X_2$, Eq.\ (\ref{eq:Pump}) simplifies to the result
(\ref{eq:PumpFirst}) quoted in the introduction. Note that,
since the parameters $X_1$ and $X_2$ are dimensionless, the
current formula contains no factor $h$, unlike the Landauer formula
(\ref{eq:Landauer}). Planck's constant may however appear in the
typical scales for the parameter dependence of the scattering matrix
$S(X_1,X_2)$.

Eq.\ (\ref{eq:Pump}) is the main result of this paper. It establishes
the link between the pumped current $I$ and the parametric derivatives
of the scattering matrix $S$. 
Several qualitative observations can already be reached on the basis of
Eq.\ (\ref{eq:Pump}).
First, for a phase coherent quantum system, the out-of-phase variation
of {\em any} pair of independent parameters will give rise to a
d.c.\ current to order $\omega$.
Second, $I$ is not quantized, unlike in the case of the electron pumps
that operate in the regime of Coulomb blockade.\cite{PLUED}
Third, if the size of the variations $\delta X_1(t) = \delta X_1
\sin(\omega t)$ and $\delta X_2(t) = \delta X_2 \sin(\omega t - \phi)$
is small compared to the characteristic correlation scales $X_{1c}$ and
$X_{2c}$ needed to change the scattering properties of the sample, we
may neglect the $X_1$ and $X_2$ dependence of the integrand in
Eq.\ (\ref{eq:Pump}), and recover the (bi)linear response formula
(\ref{eq:PumpFirst}).
On the other hand, for $\delta X_{j} \gg X_{jc}$ ($j=1,2$), the
integrand in Eq.\ (\ref{eq:Pump}) may have multiple sign changes within
the integration area $A$, and the typical value of $I$ is proportional
to $(\delta X_1 \delta X_2 X_{1c} X_{2c} \sin \phi)^{1/2}$. [Although
the typical value of the current scales as $(\sin \phi)^{1/2}$, the
$\phi$-dependence of the sample-specific current may be quite random.]

\begin{figure}
\epsfxsize=0.89\hsize
\hspace{0.04\hsize}
\epsffile{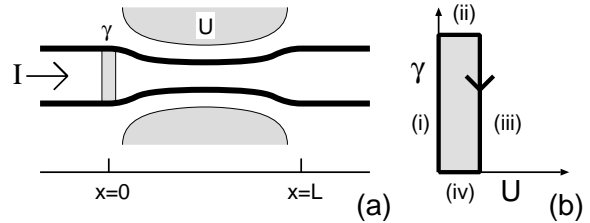}

\caption{\label{fig:3} (a) An electron pump, consisting of a 
one-dimensional wire with a tunnel
barrier at $x=0$ and an adjustable electrostatic potential $U$ for
$0 < x < L$. (b) Charge is pumped through the wire by varying the
height $\gamma$ of the tunnel barrier and the potential
$U$ in the following order: (i) $\gamma \to 
\infty$ (close barrier), (ii) $U \to \delta U$ (raise potential), (iii) $\gamma \to 0$ (open
barrier), (iv) $U \to 0$ (lower potential).} 
\end{figure}

Like the Landauer formula, the
scattering matrix formula (\ref{eq:Pump}) describes both a
classical contribution to the current and the quantummechanical
corrections. Their roles are illustrated below in two examples.
First, we consider a simple pump
in a one dimensional wire. The wire contains a tunnel barrier at $x =
0$ and for $0 < x < L$ a region where the electrostatic potential $U$ can
be varied, e.g.\ by varying the voltage of a nearby gate, see
Fig.\ \ref{fig:3}a. The Schr\"odinger Equation for this system reads
\begin{eqnarray}
  k^2 \psi(x) &=& \left( - {\partial^2 \over \partial x^2} + V(x)\right) \psi(x),
  \nonumber \\
  V(x) &=& \gamma\, \delta(x) + U\, \theta[x(L-x)],
\end{eqnarray}
where $\theta(z) = 1$ if $z > 0$ and $0$ otherwise. We pump electrons
through the system by opening and closing the tunnel barrier and
raising and lowering the potential $U$ as indicated in
Fig.\ \ref{fig:3}b. This system operates as a classical ``peristaltic''
electron pump. To find the d.c.\ current $I$,
we compute the scattering matrix $S$ and apply the scattering matrix
formula (\ref{eq:Pump}),
\begin{equation}
  I = {e L \omega \over 8 \pi^2 k}\, \delta U\, + {e \omega\, \delta U \over 16 \pi^2 k^2}  
      (\pi \sin^2 k L - \sin 2 k L).
\end{equation}
The first term is the classical contribution to the pumped current.
[Note that the local density of states for this one-dimensional system
is $1/(2 \pi k)$.] The second term is the correction due to quantum
interference.

As a second example, we consider the case where electrons are pumped
through a disordered or chaotic quantum dot. This application is
relevant for the experiments of Ref.\ \onlinecite{Marcus}. The two
(dimensionless) parameters $X_1$ and $X_2$ characterize two different
deformations of the shape of the dot, see Fig.\ \ref{fig:1}a. Unlike in
the previous example, where the pumping mechanism was of a mainly
classical origin, for a chaotic quantum dot, there is no ``classical''
contribution to the pumped current. The current results from quantum
interference and its size and direction depend on microscopic details
of the system. Pumping occurs because the wave functions near the
two point contacts are different and strongly parameter dependent,
so that different amounts of current flow through the two
contacts if the parameters $X_1$ and $X_2$ are varied.

\begin{figure}
~\vspace{-1.5cm}

  \epsfxsize=0.65\hsize
  \hspace{0.15\hsize}
  \epsffile{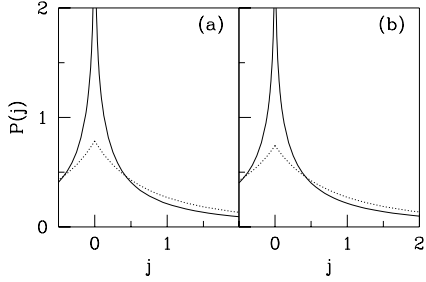}

\caption{ \label{fig:2} (a) Distribution of current $I = 
\omega j \sin \phi\, \delta X_1 \delta X_2$ for single-channel point
contacts. The case of presence (absence) of time-reversal
symmetry is shown solid (dashed). (b) The same, but with capacitive
interactions taken into account by a self-consistent Hartree 
approach, in the limit $C \ll \rho e^2$.}
\end{figure}

For a disordered or chaotic quantum dot, the statistical distribution
of the scattering matrix $S(X_1,X_2)$ and its dependence on $X_1$ and
$X_2$ are given by Random Matrix Theory.\cite{Review}  Within Random
Matrix Theory, the parameter-dependent Hamiltonian ${\cal H}(X_1,X_2)$
of the quantum dot is replaced by a large $M \times M$ hermitian matrix
$H(X_1,X_2)$,
\begin{equation} \label{eq:RMT}
  {H}(X_1,X_2) = 
     {H} + M^{-1/2} X_1 {H}_1 + M^{-1/2} X_2 {H}_2,
\end{equation}
where the matrix elements of ${H}$, ${H}_1$, and ${H}_2$
are independently and identically distributed Gaussian random numbers.
A distinction is made between the cases that time-reversal symmetry
(TRS) is present [${H}(X_1,X_2)$ is real] or absent [${H}(X_1,X_2)$ is
complex].
We now compute the distribution of the pumped current $I$ in the regime of
small parametric variations $\delta X_j \ll X_{jc}$ ($m=1,2$).\cite{Xc} 
Using the distribution of the matrices $R_{X_1}$ and $R_{X_2}$
from Ref.\ \onlinecite{BFB}, we find
that for many-channel leads ($N \gg 1$),
the current $I$ is Gaussian distributed with $\langle I \rangle = 0$ and
$\mbox{r.m.s.\ }I = e \omega \sin \phi\, \delta X_1 \delta X_2/2 \pi N$, both 
with and without TRS.
In the opposite case of single-channel leads ($N=1$), the distribution
is highly non-Gaussian, see
Fig.\ \ref{fig:2}. It has a logarithmic
singularity (cusp) at zero current in the presence (absence) of time-reversal
symmetry. The asymptotics for large $|I|$ are given by
\begin{equation}
  P(I) \propto \left\{ \begin{array}{ll} |I|^{-9/4} & \mbox{TRS}, \\
  |I|^{-3} & \mbox{no TRS}. \end{array} \right.
\end{equation}

The scattering matrix formula (\ref{eq:Pump}) allows us not only to
find the statistical distribution of the pumped current $I$, but also
the statistical correlation between $I$ and the conductance $G$ of a
chaotic quantum dot. The correlation between $I$ and $G$ shows a
remarkable dependence on the presence or absence of time-reversal
symmetry (TRS): In Ref.\ \onlinecite{BFB}, it was shown that the
statistical distribution of $R_{X_1}$ and $R_{X_2}$ is correlated with
that of the conductance in the presence of TRS only. Therefore, without
TRS, $I$ and $G$ are statistically uncorrelated for a chaotic quantum
dot, while they are correlated in the presence of TRS. In the latter
case, we find that the width of the current distribution at a fixed
conductance $G$ is proportional to $G^{1/2}$ for single-channel 
leads.\cite{foot2}

Our main result, Eq.\ (\ref{eq:Pump}), is readily extended to include
the effect of a capacitive interaction in the quantum dot within a
self-consistent Hartree treatment.\cite{footnote,BC} Following Ref.\
\onlinecite{BC}, the effect of the capacitive interaction is described
by a self-consistent electric potential $U$.  The potential $U$ is
related to the (kinetic) energy $E$ and the Fermi energy $E_F$ via $E_F
= E + U$.  Variation of $X_1$ and $X_2$ will cause a change of $U$ and
hence of $E=E_F-U$. (The Fermi energy $E_F$ is kept constant.) Hence we
have to deal with a simultaneous variation of $E$, $X_1$, and $X_2$.
These simultaneous variations can still be described by the scattering
matrix formula (\ref{eq:Pump}) provided we replace the response
matrices $R_{X_j}$ ($j=1,2$) by
\begin{eqnarray}
  R_{X_j} \to R_{X_j} + R_{E} {\partial E \over \partial {X_j}},\ \ R_{E} = -i {\partial S \over \partial E} S^{\dagger}.
\end{eqnarray}
The derivative ${\partial E / \partial {X_j}}$ reads \cite{BC}
\begin{equation}
  {\partial E \over \partial {X_j}} = - {\mbox{tr}\, R_{X_j} \over \pi C/e^2 + \mbox{tr}\, R_E}, \ \ j=1,2
\end{equation}
where $C$ is the geometrical capacitance of the dot. For
many-channel contacts, inclusion of the interactions
has no effect on the current distribution. For single-channel leads, we
have computed the current distribution for the experimentally relevant
case $C \ll e^2 \rho$, where $\rho$ is the (average) density of states
in the dot, using the distribution of the matrices $R_{X_j}$ and $R_E$
in the presence of capacitive interactions.\cite{BLFBB} The result, which
is not much different from the non-interacting case, is
shown in Fig.\ \ref{fig:2}b.

It is a pleasure to acknowledge discussions with I.\ L.\ Aleiner,
M.\ B\"uttiker, B.\ I.\ Halperin, C.\ M.\ Marcus, and Y.\ Oreg. This
work was supported by the NSF under grants no.\ DMR 94-16910, DMR
96-30064, and DMR 97-14725.


\begin{references} 

\bibitem{KJVHF} 
       L. P. Kouwenhoven {\em et al.},
       Phys. Rev. Lett. {\bf 67}, 1626 (1991).
\bibitem{PLUED} H. Pothier {\em et al.}, 
       Europhys. Lett.\ {\bf 17}, 249 (1992).
\bibitem{BS} C. Bruder and H. Schoeller, Phys. Rev. Lett. {\bf 72},
      1076 (1994).
\bibitem{SZB} B. Spivak, F. Zhou, and M. T. Beal Monod, Phys. Rev. B
      {\bf 51}, 13226  (1995).
\bibitem{SN} T. H. Stoof and Yu. V. Nazarov, Phys. Rev. B {\bf 53},
      1050 (1996); preprint (cond-mat/9707310).
\bibitem{SS} C. A. Stafford and N. S. Wingreen, Phys. Rev. Lett. 
      {\bf 76}, 1916 (1996).
\bibitem{OKKVH} T. H. Oosterkamp {\em et al.},
      Phys. Rev. Lett. {\bf 78}, 1536 (1997).
\bibitem{BBS} Ph. Brune, C. Bruder, and H. Schoeller, Phys. Rev. B
      {\bf 56}, 4730 (1997).
\bibitem{F} K. Flensberg, Phys. Rev. B {\bf 55}, 13118 (1997).
\bibitem{BP} M. H. Pedersen and M. B\"uttiker, preprint (cond-mat/9803306).
\bibitem{AA} I. L. Aleiner and A. V. Andreev, Phys.\ Rev.\ Lett.\ {\bf 81},
      1286 (1998).
\bibitem{Marcus} M. Switkes, C. M. Marcus, K. Campman,
      and A. C. Gossard (in preparation).
\bibitem{Review} For a review, see C. W. J. Beenakker, Rev. Mod. Phys. 
      {\bf 69}, 731 (1997).
\bibitem{footnote} The effect of a capacitive interaction for a similar
      electron pump in a dot with two single-channel leads is considered in
      Ref.\ \onlinecite{AA}.
\bibitem{BTP} M. B\"uttiker, H. Thomas, and A. Pr\^etre, Z. Phys. B {\bf 94}, 
133 (1994).
\bibitem{Xc} For the dimensionless parameters $X_1$ and $X_2$ defined by
      Eq.\ (\ref{eq:RMT}) $X_{1c} = X_{2c} = N^{1/2}$, see  
      Z. Pluha\u{r} {\em et al.}, 
      Phys. Rev. Lett. {\bf 73}, 2115 (1994) and
      K. Frahm, Europhys. Lett. {\bf 30}, 457 (1995).
\bibitem{BFB} P. W. Brouwer, K. M. Frahm, and C. W. J. Beenakker, Phys.
      Rev. Lett. {\bf 78}, 4737 (1997).
\bibitem{foot2} A more detailed analysis 
      shows that in the
      presence of time-reversal symmetry the quantity $I/\sqrt{G}$ is
      statistically independent from $G$.
\bibitem{BC} M. B\"uttiker, J. Phys. Condens. Matter {\bf 5}, 9361 (1993);
      M. B\"uttiker and T. Christen, in {\em Quantum Transport in
      Semiconductor Submicron Structures}, B. Kramer ed., NATO ASI
      Ser.\ E, Vol.\ 326 (Kluwer, Dordrecht, 1996).
\bibitem{BLFBB} P. W. Brouwer {\em et al.},
      Phys. Rev. Lett. {\bf 79}, 913 (1997).

\end{references}
\end{document}